\documentclass[ApJ,twocolumn]{aastex63}
\usepackage{textcomp,gensymb,graphicx}

\newcommand{\nuc}[2]{\ensuremath{^{#2}\rm{#1}}}
\newcommand{\alphap}{\ensuremath{(\alpha,\rm{p})}}

\graphicspath{{./}{figures/}}

\newcommand{\msun}{\ensuremath{M_\sun}}

\newcommand{\kbbar}{\ensuremath{\mathrm{k_B \; baryon^{-1}}}}

\newcommand{\Ye}{\ensuremath{Y_{\mathrm{e}}}}
\newcommand{\thismodel}{{\sc s18.88}}
\newcommand{\ti}{{\nuc{Ti}{44}}}

\newcommand{\MPA}{Max Planck Institute for Astrophysics, Postfach 1317, 85741 Garching, Germany}
\newcommand{\TUM}{Technical University of Munich, TUM School of Natural Sciences,
Physics Department, James-Franck-Straße 1, D-85748 Garching, Germany}

\newcommand{\vertex}{{\sc Prometheus-Vertex}}

\begin{document}

\title{Production of \nuc{Ti}{44} and iron-group nuclei in the ejecta of 3D neutrino-driven supernovae}

\correspondingauthor{Andre Sieverding}
\email{asieverd@mpa-garching.mpg.de}

\shorttitle{\nuc{Ti}{44} and iron-group nuclei in 3D supernovae}
\shortauthors{Sieverding et al.}

\author[0000-0001-8235-5910]{Andre Sieverding}
\affiliation{\MPA}
\author[0000-0003-1120-2559]{Daniel Kresse}
\affiliation{\MPA}
\affiliation{\TUM}
\author[0000-0002-0831-3330]{Hans-Thomas Janka}
\affiliation{\MPA}

\date{\today}
\begin{abstract}
The radioactive isotopes of \ti\ and \nuc{Ni}{56} are important products of explosive nucleosynthesis, which play a key role for supernova (SN) diagnostics and were detected in several nearby young SN remnants. However, most SN models based on non-rotating single stars predict yields of \ti\ that are much lower than the values inferred from observations. We present, for the first time, the nucleosynthesis yields from a self-consistent three-dimensional (3D) SN simulation of a $\sim$19~$M_\odot$ progenitor star that reaches an explosion energy comparable to that of SN~1987A and that covers the evolution of the neutrino-driven explosion until more than 7~seconds after core bounce. We find a significant enhancement of the Ti/Fe yield compared to recent spherically symmetric (1D) models and demonstrate that the long-time evolution is crucial to understand the efficient production of \ti\ due to the non-monotonic temperature and density histories of ejected mass elements. Additionally, we identify characteristic signatures of the nucleosynthesis in proton-rich ejecta, in particular high yields of $^{45}$Sc and $^{64}$Zn.
\end{abstract}

\keywords{supernovae: general; supernovae: individual (SN~1987A, Cas~A); nuclear reactions, nucleosynthesis, abundances; neutrinos; hydrodynamics}

\section{Introduction}

Since the first observations of the \nuc{Ti}{44} gamma-ray line from the core-collapse supernova (CCSN) remnant of Cassiopeia A (Cas~A) \citep{IyDiBl94} it has been clear that its ejected mass of \ti\ is at least $10^{-4}$~\msun. Similar yields have been obtained for SN~1987A \citep[see][for a recent compilation of observations]{Weinberger.Diehl.ea:2020}.
These measurements are in tension with theoretical SN calculations, in particular spherically symmetric ones \citep{WoHeWe02,WoHe07}, even when including stellar rotation \citep{Chieffi.Limongi:2017}, which predict such a large production of \nuc{Ti}{44} only for very energetic explosions and consequently in association with high \nuc{Ni}{56} masses.  
The observed, high yields of \nuc{Ti}{44} were recognized as a problem for theoretical SN models \citep{The.Diehl.ea:2006} and asymmetric explosions had been suggested before as a possible solution.
With parameterized axisymmetric (2D) and strongly deformed explosion simulations, \citet{NaHaSa98} showed that high ratios of \nuc{Ti}{44}/\nuc{Ni}{56} can be obtained from multi-dimensional explosions, where they put emphasis on high-entropy outflows along the poles of the explosion \citep[see also][]{MaNaNo02,Maeda.Nomoto:2003}.

More recently, \citet{Wongwathanarat.Janka.ea:2017} presented results of full 3D simulations that can explain the \nuc{Ti}{44} distribution and yield of Cas~A. However, these authors employed a parameterized neutrino-driven engine to obtain sufficiently energetic explosions and the models produced \nuc{Ti}{44} in sufficient amounts only under the assumption that most of the neutrino-heated ejecta had an electron fraction near $\Ye = 0.5$. Self-consistent, parameter-free models in 2D and 3D that allow to simulate for significantly longer periods than 1\,s after core bounce have become possible only recently. Such long-time simulations were performed in extensive studies in axisymmetry \citep[2D;][]{Bruenn.Sieverding.ea:2023,Vartanyan.Burrows:2023} and also in 3D 
\citep{MuMeHe17,Mueller.Tauris.ea:2019,Stockinger.Janka.ea:2020,Sandoval.Hix.ea:2021,Bollig.Yadav.ea:2021,Coleman.Burrows:2022,Burrows.Vartanyan.ea:2023,Vartanyan.Burrows.Wang.ea:2023,Pajkos.VanCamp.ea:2023}.

Several of the recent studies of 3D models also discussed aspects of nucleosynthesis, albeit with various limitations. For example, \citet{Wang.Burrows.ea:2023} discussed the nucleosynthesis of the innermost, late-time neutrino-driven ejecta of 3D CCSN models without including the material that contributes most of the \nuc{Ti}{44} and \nuc{Ni}{56}.
\citet{Stockinger.Janka.ea:2020} and \citet{Bollig.Yadav.ea:2021} discussed the nucleosynthesis of 3D models only with small reaction networks that are not suitable to reliably estimate the production of \ti\ or elements heavier than iron. \citet{Sandoval.Hix.ea:2021} used a larger, but still quite limited reaction network in a 3D simulation of a low-mass, zero-metalicity star, carried on to shock breakout.

In our study reported here, we present, for the first time, post-processing nucleosynthesis results obtained with a large reaction network for a self-consistent 3D CCSN model covering several seconds, based on a progenitor star in the mass range of the progenitor of SN~1987A.
Our work is based on the recent 3D CCSN simulation of \citet{Bollig.Yadav.ea:2021}, which modeled the explosion of an 18.88\,$M_\odot$ progenitor star (\thismodel) followed for an evolution of over 7\,s after core bounce and reaching an energy of $\sim$10$^{51}$\,erg, close to the value inferred for SN~1987A.
We find that the 3D model exhibits a significant enhancement of the Ti/Ni ratio compared to recent 1D models and compared to the same 3D model without the long-time evolution.

In Section~\ref{sec:methods}, we provide a brief summary of the simulation setup and describe the post-processing procedure for the nucleosynthesis. In Section~\ref{sec:nuc_overview}, we present the corresponding yields, with a focus on \nuc{Ti}{44} and \nuc{Ni}{56} in Section~\ref{sec:ti-ni}, their uncertainties in Section~\ref{sec:uncertainties}, and the important role of the 3D long-time evolution in Section~\ref{sec:long-time}. We draw conclusions in Section~\ref{sec:discussion}. 

\section{Methods}
\label{sec:methods}

\subsection{Supernova simulation}
\label{sec:simulation}

The SN simulation and the numerical code were described in detail by \citet{Bollig.Yadav.ea:2021} and here we only give a brief overview of the setup and simulation. The core collapse and explosion modeling was performed with the neutrino-hydrodynamics code \vertex, which employs an energy and velocity dependent, three-flavor neutrino transport (the \textsc{Vertex} module) that solves the first two moments of the Boltzmann transport equations with a variable Eddington factor closure and includes a state-of-the-art description of the neutrino interactions \citep[][see also \citealt{Fiorillo+2023}]{RaBuJa02,BuRaJa06}.

\textsc{Vertex} uses the ray-by-ray plus (RbR+) approximation to exploit the excellent parallel scaling of this approach as well as its feature of a variable Eddington closure based on the solution of 1D Boltzmann equations. \citet{Glas.Just.ea:2019} demonstrated that the overall hydrodynamic SN evolution in 3D simulations with the RbR+ approximation agrees very well with simulations using full multi-D transport, despite small-scale differences of the flow patterns and their physical conditions. This overall agreement is due to the fact that the highly time-variable and violently turbulent fluid motions in the neutrino heating region, which are often dominated by large-scale flow patterns, are not sensitively affected by small-scale, local variations of the neutrino field associated with the RbR+ transport. Instead, the flow evolution is mainly determined by neutrino effects (heating, cooling, lepton exchange) that are averaged over extended spatial regions and longer time intervals connected by the non-radial flows.

The CCSN run was started from a 3D calculation that followed the final 7~min of convection and nuclear burning in the O- and Ne-layers prior to iron-core collapse \citep{YaMuJa20}. Our post-processing does not include the nucleosynthesis during this pre-collapse phase. The 3D SN simulation seamlessly connects this 3D pre-collapse evolution into phases of core collapse, bounce, and neutrino-driven explosion. The success of the neutrino-driven mechanism was facilitated by the extra boost of violent convective overturn (``turbulence'') in the neutrino-heating layer behind the shock that was provided by the large-amplitude and large-scale perturbations in the 3D O-Ne layer falling into the stalled shock \citep{Bollig.Yadav.ea:2021}.
For matter at high densities the nuclear equation of state of \citet{LaSw91} with an incompressibility modulus of $K = 220$\,MeV (LS220) was employed.

The elaborate neutrino transport of the {\sc Vertex} module was applied for nearly 2~s after core bounce.
To continue the 3D simulation until more than 7~s after bounce this expensive neutrino transport was then replaced by a computationally less demanding, approximate neutrino treatment (termed \textsc{Nemesis} for
\textbf{N}eutrino-{\bf E}xtrapolation {\bf M}ethod for {\bf E}fficient {\bf SI}mulations of
{\bf S}upernova explosions) that describes the heating and cooling as well as lepton exchange of neutrinos with the stellar medium in and around the proto-neutron star (PNS). To this end, it employs the neutrino transport results from a 1D cooling simulation of an equally massive PNS, applying the same nuclear equation of state. The important effects of PNS convection on the evolution and neutrino emission are included with a mixing-length treatment in the 1D PNS simulation (for more details, see Appendix~E of \citealt{Stockinger.Janka.ea:2020} and \citealt{Kresse2023}).

In the transition from the simulation with full {\sc Vertex} neutrino transport to the long-time evolution with the {\sc Nemesis} scheme, no worrisome transients or artifacts in the behaviour of the fluid flows were observed \citep{Bollig.Yadav.ea:2021}, and we do not witness any artificial discontinuities in the thermodynamic histories of the tracer particles discussed below.
Despite the 1D nature of the neutrino source in the {\sc{Nemesis}} scheme, asymmetries in the matter flow, i.e., variations of density, temperature, and composition, are still taken into account for the local neutrino source and sink terms at late times. Therefore the accretion streams that
channel additional material into the neutrino heating region and lead to subsequent outflows evolve in a continuous manner between the two different neutrino treatments and do not exhibit any noticeable differences in their properties and dynamical behavior after the change of the neutrino handling. We think the reason for this insensitivity is once again the fact that the turbulent downflows react only to neutrino effects that are averaged over larger intervals in space and time.\footnote{A necessary prerequisite for this satisfactory outcome is the absence of global anisotropies in the PNS's neutrino emission, e.g., caused by asymmetric accretion or PNS convection. While the total energy luminosity exhibits large-scale anisotropy only on the level of some percent, there is a considerable dipole of the $\nu_e$ minus $\bar\nu_e$ number emission \citep{Bollig.Yadav.ea:2021} due to the so-called LESA (= Lepton Emission Self-sustained Asymmetry) phenomenon \citep[e.g.,][]{Tamborra+2014,OConnor+2018,Glas+2019,Vartanyan+2019}, which is connected to a hemispheric asymmetry of PNS convection. Potential effects of the LESA emission asymmetry on the neutrino-heated outflows at times later than $\sim$2\,s after bounce are not clarified by our study.}

A comparison of simulations with the {\sc{Nemesis}} scheme and full neutrino transport models will be presented in a forthcoming publication.

In Section \ref{sec:long-time}, we also show that the production of \nuc{Ti}{44} and \nuc{Ni}{56} at late times is only indirectly affected by the neutrino treatment when neutrino-heated matter rising outward in low-density plumes collides with dense downflows of cooler postshock matter at large radii. For all these reasons we do not expect multi-D details of the PNS's properties and its neutrino emission to be crucial for the discussed aspects of SN nucleosynthesis.

\subsection{Nucleosynthesis post-processing}
\label{sec:post-processing}

The post-processing of the nucleosynthesis is performed with the reaction network code 
XNet,\footnote{https://github.com/starkiller-astro/XNet} which is optimized for the use at large-scale computing facilities required to process large numbers of tracer particle trajectories. Other reaction network codes exist \citep[e.g.,][]{Reichert.Winteler.ea:2023,pynucastro2,Lippuner.Roberts.ea:2017,Meyer.Adams:2007}, but the methods for the numerical solution of the reaction network equations are well established and
we do not expect any significant differences depending on the particular code used. Our reaction network includes  5418 nuclear species up to $Z=85$, ranging from the proton-drip to the neutron-drip lines. This range is more than sufficient for the conditions found in the simulation, which do not allow for the $r$-process to operate. 
We include thermonuclear reaction rates and $\beta$-decay rates from the REACLIB reaction rate library \citep{CyAmFe10} as well as temperature-dependent electron-capture rates from \citet{Langanke.Martinez:2005} and \citet{FuFoNe80}.

The post-processing is based on the evolution of the temperature, density, and $Y_{\mathrm{e}}$ in fluid elements of the ejecta, extracted by placing Lagrangian tracer particles into the stored outputs of the 3D simulations and integrating their spatial positions by using the velocity field from the hydrodynamic results.
The network calculations are started for each tracer particle once the temperature according to its trajectory drops for the last time below the threshold of nuclear statistical equilibrium (NSE), for which we assume 8\,GK. This criterion is applied for each particle individually. Afterwards $Y_{\rm{e}}$ is determined by the reaction network, while temperature and density still follow the tracer particle history. 
In the network calculations, we include neutrino-induced charged-current reactions on free nucleons, using the neutrino luminosities and mean energies from the SN simulation. However, the change of $Y_{\rm{e}}$ due to these reactions is small because of a substantial abundance of $\alpha$ particles and typically low neutrino fluxes after freeze-out. These reactions, however, are important to make sure that we include possible contributions from the $\nu p$~process \citep{Froehlich.Martinez-Pinedo.ea:2006}.
We do not include $\nu$-induced reactions on nuclei for the $\nu$-process \citep{WoHaHo90}. 

The initial composition for tracer particles that do not reach NSE is taken from the original, 1D stellar evolution model at the onset of core collapse \citep[see][]{YaMuJa20}, neglecting the impact of the 3D O-burning and O-Ne shell merger on the composition, because the detailed composition including around 1000 isotopes is only available for the 1D model. For the present work, this approach is justified because we focus here on the nucleosynthesis products in the fraction of the ejecta that experiences freeze-out from NSE.
A study of the impact of the 3D progenitor evolution on the nucleosynthesis is subject of future work.

The tracer particles are placed in a spherically symmetric distribution into the final output of the \vertex\ simulation, i.e., at about 2\,s. We initially insert 460,800 particles in 64, logarithmically-spaced radial shells between 220~km and 10,000~km, representing 0.6~\msun. Each radial shell consists of 120 radial columns linearly spaced in the azimuthal direction and 60 columns distributed linearly with the cosine of the polar angle. 
388,960 of the tracer particles are classified as ejecta, i.e., they reach a positive total energy and radial velocity at the end of the simulation at 7\,s after bounce. Most of the remaining particles are accreted into the neutron star and a small fraction is still located in downflows of convective overturn motions, which do not allow for a final classification. The ejected particles represent 0.42~\msun\ of material. In this work we focus on the inner material that contributes to the nucleosynthesis of \ti\ and \nuc{Ni}{56}. Although the powerful explosion will eject most of the outer parts of the star, too, we do not include tracer particles to sample these layers, because they do not experience the conditions necessary to produce \ti\ and \nuc{Ni}{56}.

The tracer histories are then reconstructed with a backward integration approach until the onset of core-collapse. At the time of collapse the outermost particles are located at radii of up to 13,000~km. Note that this value is larger than the maximum radius of the particles at 2\,s because of the contraction during the collapse. From 2\,s until the end of the simulation at about 7\,s, the tracer particles are followed with forward integration. The equations of motion are solved by a midpoint integration method with 10~substeps, as described in \citet{Sieverding.Waldrop.ea:2023}. 
 Snapshots from the simulation are available for every 0.5~ms and the tracer particle positions and properties are stored with the same frequency. Intervals of less than 1~ms have been shown to be generally sufficient for the reconstruction of tracer particle histories for nucleosynthesis purposes \citep{Sieverding.Waldrop.ea:2023}. The time resolution is also sufficient to capture the temperature evolution relevant for nucleosynthesis \citep{Harris.Hix.ea:2017}.

For the evolution after 7\,s, the tracer particle histories are extrapolated following the approach presented in \citet{Harris.Hix.ea:2017}.
Based on the expansion timescale $\tau_{\rm{extrap}}$ derived from the smoothed evolution of the last 30~ms of the simulation, we assume 
$\rho(t)\propto \rm{exp}[-t/ \tau_{\rm{extrap}}]$
and
$T(t)\propto \rm{exp}[-t/ (3\tau_{\rm{extrap}})]$ for the evolution of density and temperature, respectively.
Once the temperature drops below 0.5\,GK, we transition to a slower, power-law decrease with 
$\rho(t)\propto t^{-2}$ and $T(t)\propto t^{-2/3}$. 
At the end of the simulation at 7\,s, almost all tracer particles that are classified as ejecta show temperatures in the range between 0.5\,GK and 1\,GK, for which reason the impact of the details of the extrapolation are not very relevant.  

In order to calculate integrated yields it is necessary to assign masses to the tracer particles. 
At 2\,s in the simulation, when the tracer particle locations are in a spherically symmetric configuration, it is possible to assign a volume for each tracer particle and, by integrating the density field over this volume, we obtain a mass for each particle.
With our set of 460,800 tracer particles this results in tracer particle masses ranging from $2\times 10^{-9}$\,\msun\ for the innermost particles to 
$6\times 10^{-6}$\,\msun\ for the outer particles in the lower O/Ne~shell. This is well within the range of mass resolution that was found by \citet{Nishimura.Takiwaki.ea:2015} to allow for converged nucleosynthesis results (see also Section \ref{sec:uncertainties}).

To illustrate the importance of the long-time evolution, we have also performed nucleosynthesis calculations in which we start the analytic extrapolation of the thermodynamic evolution described above already after the end of the \vertex\ simulations, neglecting the long-time extension of the 3D simulation from around 2\,s to 7\,s (see Section~\ref{sec:long-time}). We refer to the results from this set of calculations as {\textit{2s+extrap}} in the following.

\begin{figure}
    \centering
    \includegraphics[width=\linewidth]{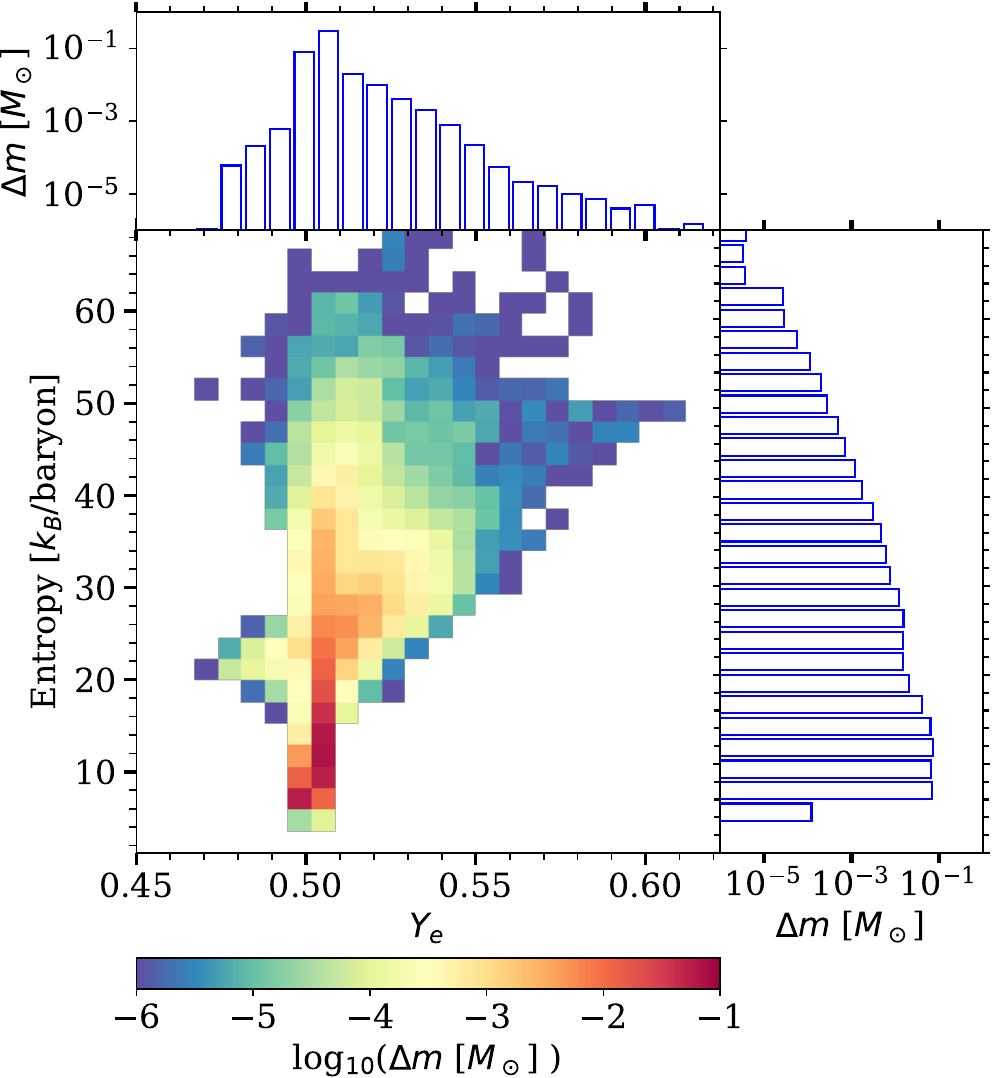}
    \caption{Distribution of the ejecta vs.\ entropy and $Y_{\rm{e}}$, evaluated for each tracer particle when it cools down to 8\,GK. }
    \label{fig:ye_ent}
\end{figure}

\citet{Navo.Reichert.ea:2022} and \citet{Harris.Hix.ea:2017} showed that there can be some differences in the yields of individual nuclear species, including \ti, 
when post-processing the nucleosynthesis compared to using extended reaction networks online with hydrodynamic simulations. These effects, however, are not well understood, are likely to be method-dependent, and applying large networks in full-physics 3D simulations is currently not feasible. For these reasons we rely on the post-processing procedure in our study.

\section{Nucleosynthesis yields}
\label{sec:yields}

We provide a brief overview of the nucleosynthesis conditions and yields in the inner ejecta in Section~\ref{sec:nuc_overview}, before we will focus in detail on the production of \ti\ and \nuc{Ni}{56} in Section~\ref{sec:ti-ni}, discuss its uncertainties in Section~\ref{sec:uncertainties} and the importance of following the hydrodynamic evolution of the ejecta over time scales of seconds in Section~\ref{sec:long-time}. 

\subsection{Overview}
\label{sec:nuc_overview}

The entropy and $Y_{\rm{e}}$ at the time of freeze-out from NSE are among the key parameters that determine the nucleosynthesis outcome for a given tracer particle.
Our tracer particles cover initial radii up to 13,000~km and include a small part of the shock-heated matter of the O/Ne shell. This material is visible as a prominent mass concentration in
Figure~\ref{fig:ye_ent}, which shows the distribution of the ejecta mass versus entropy and $Y_\mathrm{e}$ at the time when the nuclear composition freezes out from NSE. The shock-heated ejecta lead to the large accumulation of mass at $Y_\mathrm{e}\approx 0.5$ with entropies of 8--20\,\kbbar. 
\citet{Bollig.Yadav.ea:2021} constructed the mass distribution as a function of $Y_\mathrm{e}$ for matter that flows out through a sphere of fixed radius of 250~km and becomes gravitationally unbound (see their Figure~8). This material is also represented by a large fraction of our tracer particles. Therefore, we find a similar distribution as function of $Y_\mathrm{e}$. In particular, we witness predominantly proton-rich ejecta with $Y_\mathrm{e}\geq 0.5$ and values up to 0.62. There is only a small neutron-rich component with $Y_\mathrm{e}$ down to 0.47. 

\begin{figure}
    \centering
    \includegraphics[width=\linewidth]{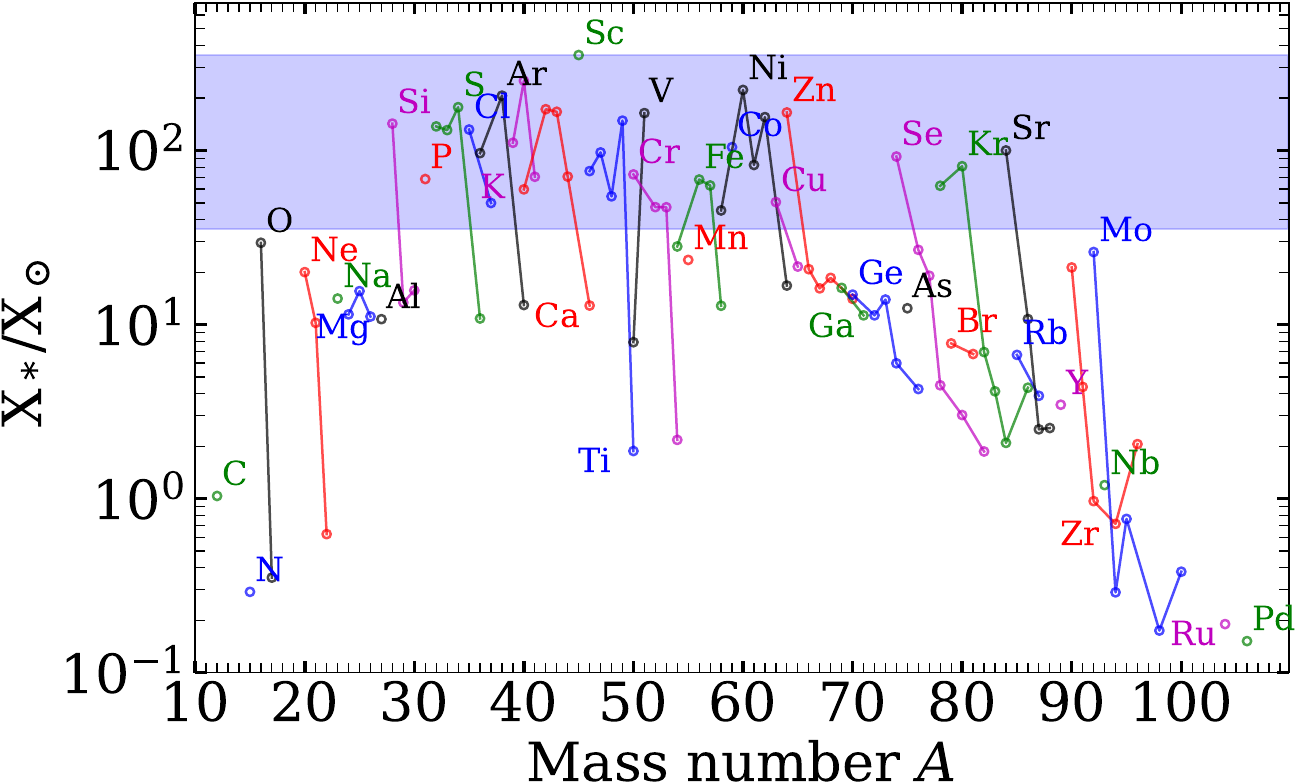}
    \caption{Production factors of stable isotopes after radioactive decay relative to the solar abundances from \citet{Asplund.Grevesse.ea:2009}. The isotope \nuc{Sc}{45} is co-produced with Fe-group elements. }
    \label{fig:pf}
\end{figure}

A similar dominance of proton-rich neutrino-heated ejecta was also seen in several other 3D SN simulations \citep{Wang.Burrows.ea:2023,Sieverding.Mueller.ea:2020,Glas.Just.ea:2019}. We also find fairly large entropies with values of up to about 65\,\kbbar. Some of the high-entropy material with $\sim$\,50\,\kbbar\ is associated with the highest values of $Y_\mathrm{e}$, because neutrino absorption reactions increase both quantities \citep{WaMuJa18}. 
The highest values of more than 60\,\kbbar, however, are reached by matter that has $Y_{\mathrm{e}}\approx 0.5$. This material provides optimal conditions for efficient production of \ti\ to be discussed in Section~\ref{sec:ti-ni}.

The integrated yields\footnote{Full yield data are available in our Core-Collapse Supernova Data Archive (\url{https://wwwmpa.mpa-garching.mpg.de/ccsnarchive/})}
of stable isotopes from the inner ejecta of our CCSN simulation of the progenitor model \thismodel\ 
are shown in Figure~\ref{fig:pf} in terms of their production factors, i.e., of their mass fractions relative to the ejecta mass represented by our tracer particles ($\sim$0.42\,\msun)
divided by the solar abundances from \citet{Asplund.Grevesse.ea:2009}. The shaded area indicates a region of a factor 10 relative to the maximum production factor, which is reached by \nuc{Sc}{45}. 
This is in good agreement with the results of \citet{Sieverding.Mueller.ea:2020}, who found a similar enhancement of \nuc{Sc}{45} in the post-processing of a 3D~simulation applying a correction to the neutrino spectra. \nuc{Sc}{45} is a key product of the $\alpha$-rich freeze-out in moderately proton-rich material \citep{FrHaLi05}.

We also find \nuc{Ni}{60} as the dominant isotope of the~$Z=28$ isotopic chain, which is co-produced with \nuc{Sc}{45} and \nuc{Zn}{64}. The predominantly proton-rich ejecta (see Figure~\ref{fig:ye_ent}) also lead to the co-production of \nuc{Zn}{64} and the $p$-nuclei of \nuc{Se}{74}, \nuc{Kr}{78}, and \nuc{Sr}{84}.
The small neutron-rich component of the ejecta (Figure~\ref{fig:ye_ent}) is responsible for the production of some \nuc{Zr}{90} and \nuc{Mo}{92} \citep{Woosley.Hoffman.ea:1992}. The final production factors of these isotopes are, however, more than one order of magnitude lower than the large production factor of \nuc{Sc}{45}. The material is not neutron-rich enough to produce any noticeable amount of \nuc{Ca}{48}.

With a single SN model it is not possible to draw relevant conclusions for chemical evolution. However, the \thismodel\ model is in agreement with the general trend \citep{WaMuJa18} that SNe of higher-mass stars contribute predominantly to the nucleosynthesis yields associated with proton-rich conditions, whereas explosions of low-mass progenitors produce more neutron-rich ejecta. We do not find any significant creation of isotopes with $A > 84$ in the material covered by our post-processing. 
However, we do not include all of the O/Ne regions that may contribute to the $p$-nuclei and other isotopes via the $\gamma$ process \citep{RoPiPs23}. 
Overall, the abundance pattern we find is similar to that from the 2D model s11 of \citet{WaMuJa18}, but with a slightly larger neutron-rich component in our case.

Most of the O-rich and C-rich layers are not covered by the tracer particles included in our analysis. Hence, \nuc{O}{16} and \nuc{C}{12} show only low production factors in Figure~\ref{fig:pf}.  

\begin{figure}
    \centering
    \includegraphics[width=\linewidth]{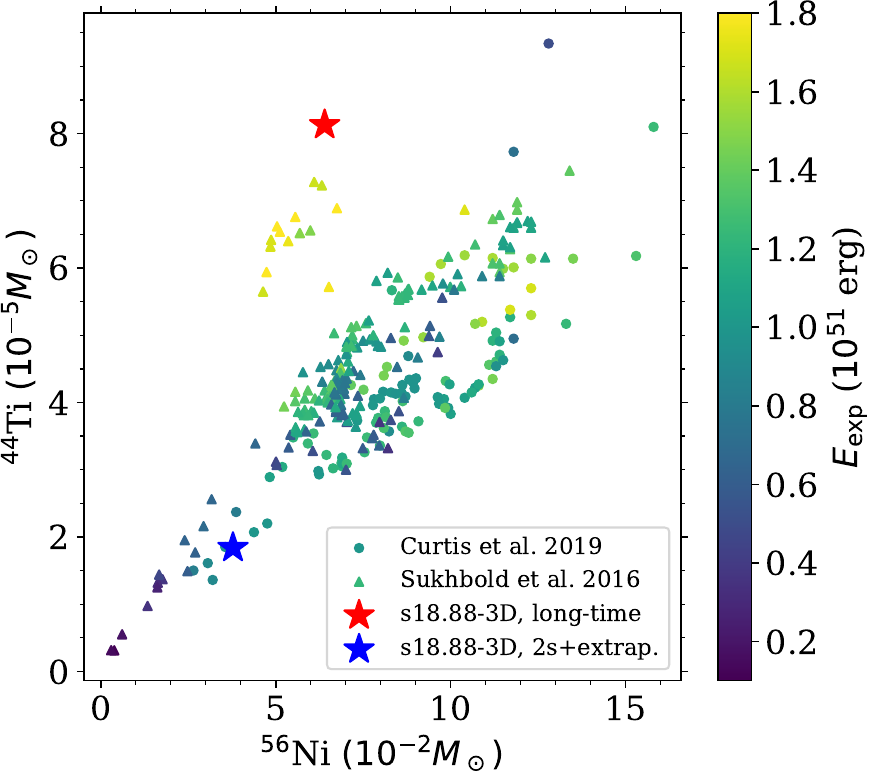}
    \caption{Yields of \nuc{Ti}{44} and \nuc{Ni}{56} obtained in this study (red star) compared to examples of 1D calculations from \citet{SuErWo16} for their calibrations N20+Z9.6 and W18+Z9.6 and from \citet{CuEbFr19s} for their ``s" and ``w" progenitor sets. There is a clear enhancement of \nuc{Ti}{44} for the same \nuc{Ni}{56} mass due to the characteristics of the 3D model. The blue star indicates the result we obtain from the extrapolation of the tracer particles after the first $\sim$2\,s of the simulation (see Section \ref{sec:long-time}).}
    \label{fig:context}
\end{figure}

\subsection{\texorpdfstring{\nuc{Ti}{44}}{44Ti} and \texorpdfstring{\nuc{Ni}{56}}{56Ni}}
\label{sec:ti-ni}

Two key observables of SN explosions and young SN remnants are the yields of \nuc{Ni}{56} and \nuc{Ti}{44}, both of which are radioactive and their decay is accompanied by the emission of characteristic radiation. In the following, we analyse the production of these two isotopes in the \thismodel\ explosion model and show how the late-time evolution of the ejecta significantly enhances the production of \nuc{Ti}{44}.

We obtain a total ejected mass of \nuc{Ni}{56} of 0.064\,\msun. This is within the range from 0.027\,\msun\ to 0.087\,\msun\ estimated by \citet{Bollig.Yadav.ea:2021} based on the small reaction network used to follow the composition during the \vertex\ simulation on the one hand, and based on the upper limit from the extended simulation assuming NSE on the other hand.

The total mass of \nuc{Ti}{44} produced by our 3D SN model is $8.13\times 10^{-5}\,\msun$. Relative to the mass of \nuc{Ni}{56}, which is known to correlate to some extent with the explosion energy \citep{SuErWo16,Ebinger.Curtis.ea:2019}, this yield of \ti\ is large compared to the results of the most recent 1D models that use parameterized explosions motivated by the neutrino-driven mechanism. Figure~\ref{fig:context} contrasts our results for \ti\ and \nuc{Ni}{56} with those of \citet{CuEbFr19s} and \citet{SuErWo16}, for which we have collected results from different available explosion calibrations and progenitor sets. 
Figure~\ref{fig:context} shows that most of the results from these 1D models fall within a relatively narrow band that indicates an approximately linear correlation between \ti\ and \nuc{Ni}{56} yields with a significant spread.

Our yields based on the long-time simulation, displayed by the red star, are clearly enhanced in \ti\ relative to \nuc{Ni}{56}. 
In Figure~\ref{fig:context}, we also show the results ($\sim$\,$0.038\,M_\odot$ of $^{56}$Ni and $\sim$\,$1.8\times 10^{-5}\,M_\odot$ of $^{44}$Ti) based on the same simulation but without the long-time extension, i.e., using only the first 2\,s from the hydrodynamic calculation and extrapolating the particle trajectories after that (blue star in Figure~\ref{fig:context}).
In this {\textit{2s+extrap}} case, the yields are in line with the results from the 1D models. The detailed reasons for the significant difference are discussed in Section~\ref{sec:long-time}.

A few models of \citet{SuErWo16} come close to the high \ti\ yield obtained with our \thismodel\ long-time simulation. However, as the color in Figure~\ref{fig:context} indicates, these models correspond to much more energetic explosions with energies of (1.6--1.8)$\times 10^{51}$\,erg. In 2D explosion simulations of a stellar progenitor of 17\,\msun, \citet{Eichler.Nakamura.ea:2018} obtained very small amounts of unbound \ti\ of less than $1.35 \times 10^{-5}$\,\msun, which is, however, probably an artifact of the deficiencies of 2D simulations.  

The \nuc{Ni}{56} produced by SN~1987A is well constrained from the light-curve tail to be around 0.075\,$M_\odot$ \citep[see, e.g.,][]{Bouchet.Phillips.ea:1991,Utrobin+2021}, but the \ti\ yield inferred from observations of the remnant of SN~1987A is subject to large uncertainties. While our yield of \nuc{Ti}{44} is somewhat higher than the estimate of $(5.5 \pm 1.7) \times 10^{-5}\,M_\odot$ by \citet{SeTiMa14} based on fitting the late light curve of SN~1987A, it is outside of the error bars reported by other studies, which commonly found a range of (1--2)$\times 10^{-4}$~\msun\ \citep{Larsson.Fransson.ea:2011,JeFrKo11,GrLuTs12,Boggs.Harrison.ea:2015}. However, the explosion energy of SN~1987A was inferred from light-curve modeling to be around (1.3--1.5)$\times 10^{51}$~erg \citep[e.g.,][]{Ar87,Utrobin2005,UtWoJa15,Utrobin+2019,Utrobin+2021}, which is 20--50\% higher than the energy of our 3D SN model. 
Assuming a linear scaling of \ti\ and \nuc{Ni}{56} with the explosion energy, we would expect (1.06--1.22)$\times 10^{-4}$~\msun\ of \ti\ for the range of explosion energies estimated for SN~1987A. Such higher values would be accompanied by an increased mass of \nuc{Ni}{56}, which in our 3D SN model is still somewhat less than the value of $\sim$\,0.075\,\msun\ of SN~1987A.

\begin{figure}
    \centering
    \includegraphics[width=\linewidth]{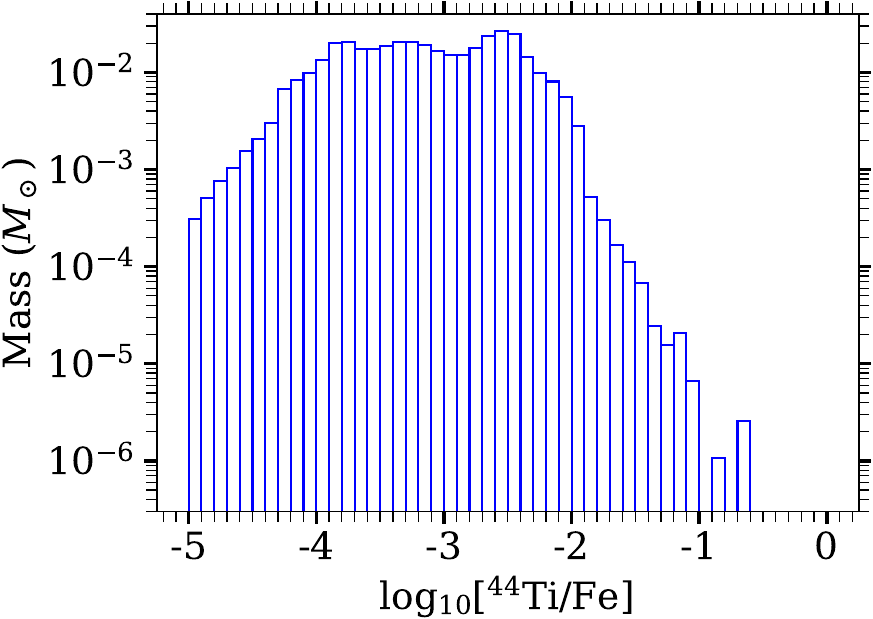}
    \caption{Histogram of the distribution of the ejecta mass as a function of the Ti/Fe ratio. Here, ``Fe" represents the Fe-group, including all isotopes of Fe, Co, and Ni.}
    \label{fig:ti-ni_hist}
\end{figure}

Cas~A is another nearby, young SN remnant. The ejected amount of \ti\ is better determined for Cas~A than for SN~1987A. The total mass is estimated to be $(1.37\pm 0.19)\times 10^{-4}$~\msun\  based on an analysis of the transitions associated with the decay of \ti\ observed with the SPI spectrometer on INTEGRAL \citep{Siegert.Diehl.ea:2015}. A similar value of $1.54 \pm 0.21 \times 10^{-4}$~\msun\ was obtained by \citet{Grefenstette.Fryer.ea:2017} based on NuSTAR high-energy X-ray observations. Several other groups have also reported similar results from their analyses \citep{TsKrLu16,Wang.Li.ea:2016}. 

The energy of the SN producing the Cas~A remnant is expected to have been rather high \citep{Laming+2003,Hwang+2003,HwLa12} and was estimated by forward and reverse shock modeling to be (2--2.3)$\times 10^{51}$\,erg \citep{Orlando+2016}. Again assuming that the \ti\ mass scales with the SN energy, the observed mass of \ti\ is also in reach of our 3D model.
 
For Cas~A, the spatial distribution of Fe and \ti\ was determined by \citet{Grefenstette.Fryer.ea:2017} and suggests the existence of locally Fe-rich ejecta components that are not enriched in \ti\ to a detectable level. There is also some evidence for the presence of \ti\ in Fe-poor regions. These observations put most nucleosynthesis models to a test, because Ti and Fe are usually co-produced in an $\alpha$-rich freeze-out. Figure~\ref{fig:ti-ni_hist} shows the mass histogram of the Ti/Fe ratios from our tracer particles, which we compute as the ratio of the \ti\ mass fraction and the summed mass fractions of all isotopes in the Fe-group, which is mostly dominated by \nuc{Ni}{56}. We do not find any tracer particles that show $X$($^{44}$Ti)$\,>\,X$(Fe-group), but there is a small component with a strong enrichment of \ti\ reaching \ti/Fe mass ratios of $\sim$\,10\% and more. These high values are the result of tracer particles with exceptionally high mass fractions of \ti, rather than because of low mass fractions of Fe-group nuclei. 

Judged on the basis of our results, observational hints to the presence of regions containing \ti\ without any visible Fe in the Cas~A supernova remnant \citep{Grefenstette.Fryer.ea:2017} should evidence the existence of iron that has not yet been heated by the reverse shock and thus cannot be observed.  
The existence of a mechanism, possibly constrained to rare conditions, that leads to $X$($^{44}$Ti)$\,>\,X$(Fe-group), either because of special nucleosynthesis or potentially connected to late-time mixing effects, cannot be excluded, however.
On the other hand, the distribution in Figure~\ref{fig:ti-ni_hist} extends to very low values of Ti/Fe, indicating the presence of material that is rich in Fe, but deficient in \ti. This is mostly material that undergoes complete Si-burning and material with $Y_\mathrm{e}<0.5$, both of which suppresses the production of \ti\ \citep{MaTiHu10}. 

These findings suggest compatibility in principle with the observations by \citet{Grefenstette.Fryer.ea:2017}, but a detailed comparison with the spatial distribution of Ti and Fe in Cas~A needs to be deferred to future work, because it requires a further continuation of our SN simulation towards the remnant phase. However, our current results demonstrate for the first time that self-consistent 3D simulations of neutrino-driven CCSNe are basically consistent with the observed amounts of \ti\ in the remnants of SN~1987A and Cas~A. 

\subsection{Model uncertainties}
\label{sec:uncertainties}

The yield of \ti\ is well resolved in terms of the number of tracer particles. A number of at least 225 particles accounts for 10\% of the total \ti\ yield and 5732 tracer particles represent 50\% of the yield. To exclude any sensitivity to the placement of these particles, we have repeated the calculation with an alternative set of 102,400 tracer particles placed in a different configuration and covering a slightly larger part of the simulation domain from 300~km out to 30,000~km in 32 radial shells. With this second set of tracer particles, we find practically the same yields of \ti\ and \nuc{Ni}{56} to within less than 5\%. 
 
In our simulation, the vast majority of \ti\ is produced in the neutrino-heated ejecta, which reach peak temperatures in excess of 10\,GK and undergo an $\alpha$-rich freeze-out from NSE. The contribution from explosive burning in the O-shell is negligible. 
 \citet{Chipps.Adsley.ea:2020} re-evaluated theoretical and experimental constraints on the \nuc{Ti}{44}\alphap\nuc{V}{47} reaction rate, which is important for the production of \nuc{Ti}{44} in CCSNe. Their recommended rate is similar to the reaction rate in the REACLIB database used for our study. We performed calculations with the tracer particle trajectories that produce the highest mass fractions of \nuc{Ti}{44} and found that the final mass fraction of \nuc{Ti}{44} is only reduced by less than 5\% with the reaction rate from \citet{Chipps.Adsley.ea:2020} compared to the current REACLIB rate. A complete analysis of the sensitivity of the  \ti\ production to nuclear reaction rate uncertainties is beyond the scope of the present work. However, it is important to note that the late-time contribution to the \ti\ yield stems from lower temperatures and densities than commonly assumed in sensitivity studies \citep{MaTiHu10,Hermansen.Couch.ea:2020}. This might affect the conclusions drawn from those previous studies. 

 \begin{figure}
     \centering
     \includegraphics[width=\linewidth]{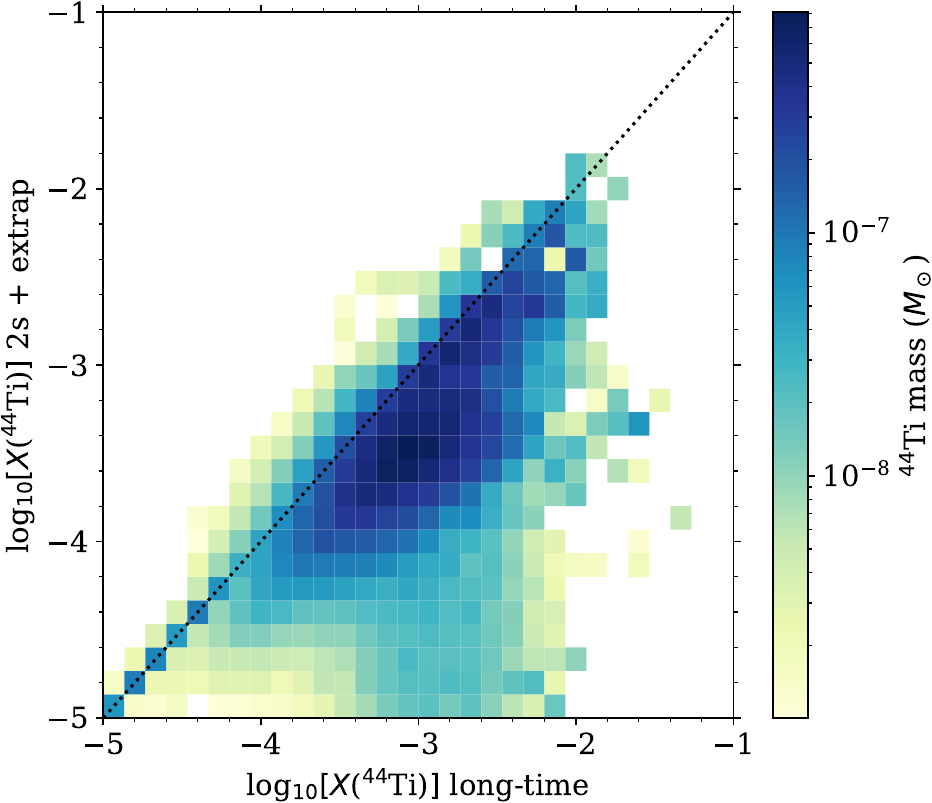}
     \caption{Distribution of the \ti\ mass fraction in matter that is ejected until 2\,s. Nucleosynthesis calculations extrapolating the tracer particle evolution after 2\,s are compared to the results based on the long-time hydrodynamic evolution of the same ejected particles until 7\,s. Contributions from tracer particles that give the same results in both cases are on the diagonal, which is indicated as black dotted line. The visible shift towards the lower right reflects the trend of higher mass fractions of \ti\ in the long-time simulation.}
     \label{fig:extrap_hist}
 \end{figure}
 
\subsection{Effect of the long-time evolution}
\label{sec:long-time}

We identify two main aspects that enhance the \ti\ yield found by us relative to the values expected from typical 1D models and previous multi-dimensional models: 
 (1) the change of the total ejecta mass that continues to grow from 2\,s to 7\,s, because neutrino-heated, high-entropy gas is still ejected from the deep SN core at late times; and (2) a more complex temperature and density evolution due to episodic compression and heating in collisions with secondary shocks, which involves non-monotonic behavior and, in general, slower cooling and expansion compared to the extrapolation. In order to illustrate these effects in the following, we evaluate the nucleosynthesis in our {\textit{2s+extrap}} calculation by only taking the first 2\,s of the hydrodynamic evolution and starting the extrapolation described in Section \ref{sec:methods} of the thermodynamic histories of ejected tracers already at this point. 
\subsubsection{Ejecta mass and trajectory extrapolation}
\label{sec:ejecta_mass}

\begin{figure}
    \centering
    \includegraphics[width=\linewidth]{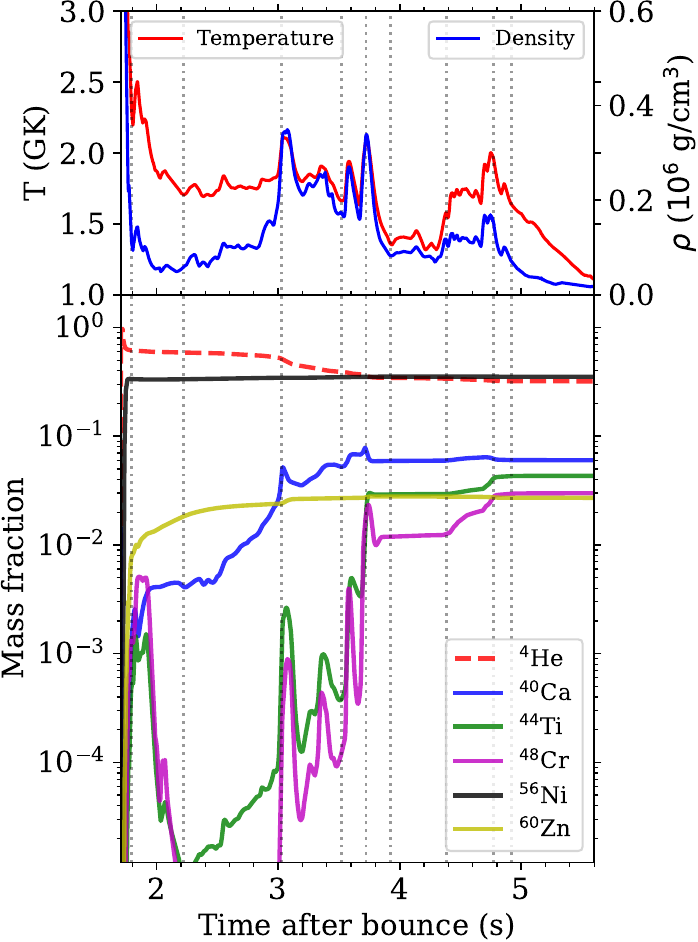}
    \caption{Composition and thermodynamic evolution of the blue tracer particle shown in Figure~\ref{fig:contour}, which produces a high mass fraction of \ti. The dotted vertical lines indicate the times picked for the snapshots in Figure~\ref{fig:contour}.
    The top panel displays the temperature and density history of the particle. The bottom panel shows the corresponding evolution of the mass fractions of several isotopes as obtained by the post-processing of this trajectory. The non-monotonic temperature evolution, in particular the temperature peaks at around 3.7\,s and 4.8\,s, increases the \ti\ mass fraction.}
    \label{fig:ti44_example}
\end{figure}

While the ejecta mass covered by the tracer particles is 0.42\,\msun\ with the full evolution, only 0.33\,\msun\ can already be classified as ejecta at 2\,s. The matter that is responsible for this difference is mostly located at radii below 1000\,km with mostly moderately negative radial velocities at 2\,s. This is the kind of material that is likely to experience multiple reheating episodes by secondary shocks later on, leading to complicated dynamics and thermal evolution that permit very efficient production of \ti. 

Therefore, these late ejecta with a mass of 0.09\,\msun\ contribute $3.8\times 10^{-5}$\,\msun\ of \ti, which accounts already for more than half of the \ti\ produced in total by the long-time model in addition to the $\sim$\,$1.8\times 10^{-5}\,M_\odot$ of the {\textit{2s+extrap}} calculation. The rest of the difference in the \ti\ yield is explained by the fact that the real hydrodynamic evolution enables the particles that are identified as ejecta already at 2\,s to create considerably more \nuc{Ti}{44} than with the extrapolated histories, namely $\sim$\,$4.3\times 10^{-5}\,M_\odot$ instead of $\sim$\,$1.8\times 10^{-5}\,M_\odot$.

This means that the increase of the ejecta mass by 0.09\,\msun\ or 27\%  
enhances the \ti\ yield by nearly a factor of 2 from $\sim$\,$4.3\times 10^{-5}\,M_\odot$ to $8.13\times 10^{-5}\,M_\odot$ in total. The boost is even more than a factor of 4 when the \ti\ mass of the {\textit{2s+extrap}} calculation is considered, because the late-time hydrodynamic evolution of the 3D SN model is very efficient at producing \ti. In contrast, the yield of \nuc{Ni}{56} grows much more moderately from 0.038\,\msun\ in the {\textit{2s+extrap}} case to 0.064\,\msun\ for the long-time model. This increase is only slightly over-proportional compared to the growth of the total ejecta mass. The correspondingly steep increase of \ti/\nuc{Ni}{56} mass ratio is also illustrated by the two stars in Figure~\ref{fig:context}. 

Figure~\ref{fig:extrap_hist} displays the correlation between the mass of \ti\ made by the long-time simulation and the {\textit{2s+extrap}} calculation that applies the extrapolation after 2\,s. This only includes the tracer particles that can be identified as part of the ejecta already at 2\,s. In both of the calculations we find high mass fractions of \ti, namely up to $\sim$0.05 and $\sim$0.013, respectively, and Figure~\ref{fig:extrap_hist} shows that the bulk of the \ti\ yield comes from particles with mass fractions around $10^{-3}$. However, it also illustrates a clear trend of the long-time simulation to increase the \ti\ production of the bulk by shifting the dark blue region towards significantly higher mass fractions. 

\begin{figure*}
    \centering
    \includegraphics[width=\textwidth]{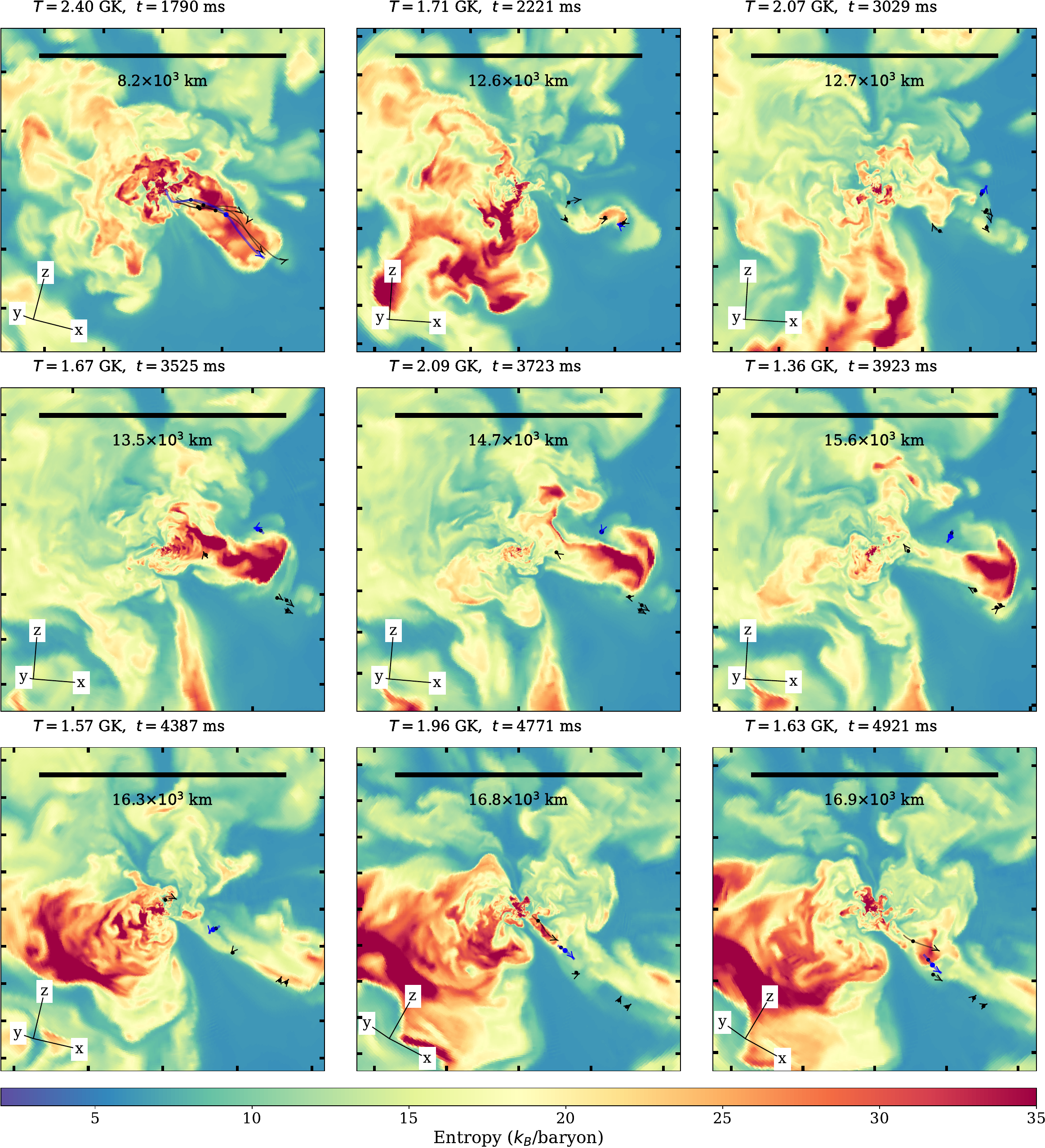}
    \caption{Time series of 2D slices of the entropy in the 3D SN model. Markers indicate the current positions of a number of tracer particles and lines illustrate the projection of the path of the particles onto the plane of the slice within 200\,ms around the time of the snapshot. One particle that very efficiently produces \ti\ (see Figure~\ref{fig:ti44_example}) is highlighted in blue. The temperature at the location of the highlighted particle as well as the time after core bounce are shown above each panel. The times of the snapshots are also indicated in Figure~\ref{fig:ti44_example}. Only the highlighted particle is always located in the plane of the slice, which is rotated slightly from the initial orientation, as indicated by the tripod in each panel. The other particles that are shown in black are projected onto this slice. In the last panel, all particles are within 1000~km of the plane.}
    \label{fig:contour}
\end{figure*}

\subsubsection{Non-monotonic trajectories}
\label{sec:nonmontraj}

For the case of exemplary tracer particles, Figures~\ref{fig:ti44_example} and~\ref{fig:contour} visualize how the violent flow dynamics of the 3D SN model can lead to complicated, often non-monotonic temperature histories of the tracers. We will now describe why these temperature non-monotonicities facilitate enhanced production of \ti. 

\citet{MaTiHu10} demonstrated that parameterizations of homologous expansion or similar descriptions lead to mass fractions of \ti\ of the order of $10^{-5}$--$10^{-4}$ when exploring the parameter space of peak temperature and density. On the other side, our non-monotonic temperature histories allow for \ti\ mass fractions of up to around 0.1. 
The general effects in such trajectories were already discussed by \citet{MaTiHu11}, based on parameterizations intended to model reverse shocks and multi-dimensional asymmetries. 
With monotonic trajectories, the final abundances are determined by a hierarchy of initially global and progressively local equilibria of nuclear reactions and a single phase of non-equilibrium evolution. Non-monotonic trajectories, in contrast, can experience multiple transitions between local equilibria and  non-equilibrium phases \citep{MaTiHu11}.

Figure~\ref{fig:ti44_example} shows an example for the evolution of the mass fractions for a non-monotonic trajectory. Initially, the particle freezes out from NSE at about 1.8\,s after core-bounce with $Y_{\rm{e}}\approx 0.51$, leading to large abundances of $\alpha$ particles and \nuc{Ni}{56} and some free protons.  At this stage, we can indeed confirm \ti\ mass fractions between $10^{-5}$ and $10^{-4}$ as reported by \citet{MaTiHu10}. As the particle remains at a temperature around 2\,GK for more than 1\,s, \nuc{He}{4} is gradually converted to heavier isotopes, slowly increasing the mass fraction of \ti.
Furthermore, \nuc{Ca}{40}, which is also shown in the bottom panel of Figure~\ref{fig:ti44_example}, is produced with a significant mass fraction and, due to the proton-excess, also \nuc{Cr}{46,47} and \nuc{Fe}{52} reach mass fractions larger than $10^{-2}$. \nuc{Zn}{60}, which eventually decays to \nuc{Ni}{60}, also reaches a mass fraction of almost 0.03, but \nuc{C}{12}, \nuc{O}{16}, \nuc{Mg}{24}, and \nuc{Si}{28} remain below $10^{-3}$, indicating an efficient flow to the $Z\geq 20$ region.\footnote{The energy feedback from this recombination of $\alpha$ particles at $T < 2.5$\,GK is not taken into account in this study and may contribute to differences between in-situ networks \citep{Harris.Hix.ea:2017,Navo.Reichert.ea:2022} and post-processing nucleosynthesis.}

When the temperature drops from over 2\,GK to about 1.5\,GK at about 3.8\,s after bounce, \ti\ is formed with a mass fraction of about $3\times 10^{-2}$ in a freeze-out from local equilibria. \nuc{Cr}{48} is also produced at the same time.
At this point, the particle is still not finally ejected, but it gets heated again to a temperature slightly above 2\,GK at $\sim$4.7\,s. 
At this stage, however, the density is rather low and, in spite of the high temperature, \ti\ is not destroyed, but instead its abundance is increased, mostly by $\alpha$ captures on \nuc{Ca}{40}, which is slightly reduced in the process. It is only after more than 5\,s after bounce, that the particle is eventually ejected and the temperature decreases monotonically afterwards, and charged-particle reactions come to a halt.

The vertical dotted lines in Figure~\ref{fig:ti44_example} indicate the times of the snapshots shown in Figure~\ref{fig:contour}, which displays the spatial position and trajectory of the exemplary tracer particle highlighted in blue and superimposed on the hydrodynamic background.
Besides this particle, the locations of several other particles in its vicinity are also shown in black. Thin lines mark the path of the particles from 200~ms before until 200~ms after each snapshot. The length of the lines thus gives an indication of the particles' velocities.

The first two panels of Figure~\ref{fig:contour} show the initial ejection of the tracer particle, driven by a high-entropy, neutrino-heated plume. The particle initially moves with a high velocity and the expansion allows the particle to cool to a temperature of $\sim$1.7\,GK at 2.2\,s. However, the high-entropy plume dissipates by shear interaction with surrounding downflow material and partly mixes with the lower-entropy and higher-density gas in the downflow. Therefore, the radial velocity of the tracer particle becomes negative again and the temperature increases due to compression, as seen in the third snapshot in Figure~\ref{fig:contour}.
Afterwards, until almost 5\,s after bounce, the particle remains at a relatively stable radius and temperature, with some significant fluctuations driven by the interaction with nearby material and secondary shocks.

For example, the middle row in Figure~\ref{fig:contour} shows another high-entropy plume emerging from the close vicinity of the PNS. This outflow does not directly hit the fluid element represented by the tracer particle under consideration, but the close passage of this outflow still leads to (shock) compression phases and, therefore, causes the peaks in the density and temperature history between 3.5\,s and 3.8\,s in Figure~\ref{fig:ti44_example}. Because the rising plume creates flow vorticity, the plasma at the location of the particle is swept into cavities where the gas expands. For this reason, the temperature and density at the particle's position drop, which is visible by the lower-temperature phase between 3.8\,s and 4.3\,s. The third panel of the middle row marks such a temperature minimum.
The first panel of the bottom row of Figure~\ref{fig:contour} shows that the particle continues to move slowly inward and the fluid element gets compressed and reheated to about 1.8\,GK at 4.4~s. 

The middle panel of the bottom row shows that, at about 4.8\,s after bounce, another fast plume rises outward from near the PNS. This time, the plume collides with the particle and drags it along, increasing its temperature by shock-heating to almost 2\,GK. This leads to a final growth of the \ti\ mass fraction and the particle's ultimate ejection with a velocity of a few 1000~km/s, as shown in the last panel of the bottom row.  Figure~\ref{fig:ti44_example} shows that in the last reheating phase between 4.3\,s and 4.8\,s the density is significantly lower than the density during the previous high-temperature phases, reflecting the decreasing density in the gradually expanding high-entropy bubbles, which are pushed from below by newly expelled neutrino-driven outflows. 

It is important to note that the dynamics of this tracer particle are not directly determined by neutrino interactions and heating, but by the hydrodynamic interaction with the neutrino-driven outflows that continuously emerge from deeper inside and collide with neutrino-heated gas expelled before and with lower-entropy downflow material swept up by the SN shock. This is important for the nucleosynthesis, because there is no major impact of neutrino interactions on $Y_\mathrm{e}$ in this particle. Hence, $Y_\mathrm{e}$ remains close to 0.5, which is ideal for the production of \ti.

The final mass fraction of \ti\ then depends decisively on the densities and temperatures reached in the secondary peaks. It also depends on the composition that results from the initial freeze-out and subsequently on the number of reheating episodes. Ejected fluid elements that undergo complete Si-burning under high-density conditions or that experience freeze out from NSE with less $\alpha$-particles do not contribute much to the synthesis of \ti. Similarly, particles with secondary temperature peaks exceeding 3\,GK do not contribute either, because much of the \ti\ is converted into \nuc{Ni}{56} if the temperatures are too high. These rather complicated dependencies do not allow to predict the final \ti\ production based on simple characteristics such as the peak temperature and density of any given trajectory.

\section{Conclusions}
\label{sec:discussion}

We have presented the results from nucleosynthetic post-processing of the innermost 0.42~\msun\ of the ejecta from a self-consistent 3D simulation of the neutrino-driven SN explosion of a $\sim$19~\msun\ progenitor \citep{Bollig.Yadav.ea:2021}. The explosion model had been followed for $\sim$7\,s after bounce until the explosion energy saturated at about $10^{51}$\,erg. Our detailed calculations with a large nuclear reaction network show the characteristic signatures of the nucleosynthesis in predominantly proton-rich outflows, with a strong enhancement of \nuc{Sc}{45} and \nuc{Zn}{64}. The small neutron-rich ejecta component of this SN model does not noticeably affect the nucleosynthesis pattern.

We found that 0.064\,\msun\ of \nuc{Ni}{56} and $8.13\times 10^{-5}$\,\msun\ of \ti\ are ejected. The mass of \nuc{Ni}{56} is in agreement with estimates from the original explosion simulation of \citet{Bollig.Yadav.ea:2021}, and it is also compatible with the value inferred for SN~1987A, in particular after proper rescaling with the explosion energy. The same conclusion applies for the measured masses of \ti\ in the remnants of SN~1987A and Cas~A. 

We found that, compared to parameterized 1D explosions, the production of \ti\ is strongly enhanced in our 3D SN model. This difference can be explained, to a large part, by the long-time evolution that is covered by the explosion simulation. On timescales of several seconds after core bounce, additional material that efficiently produces \ti\ is ejected, and the non-monotonicity of the density and temperature histories of the tracer particles at late times also enhances the production of \ti. 

Our results suggest that long-time 3D simulations of neutrino-driven explosions are able to reproduce the observed signatures of radioactive nuclei produced in CCSNe of massive stars and that such simulations are necessary to fully understand the nucleosynthesis of important radioactive isotopes such as \ti. Most studies of the sensitivity of the \ti\ formation to nuclear physics uncertainties, stellar structure, or explosion properties were carried out based on monotonic parameterizations of the particle trajectories or 1D hydrodynamic simulations. At this point, it is unclear if the importance of the late-time mass-ejection dynamics in 3D explosions, which involves lower and non-monotonically evolving temperatures due to a highly turbulent, multi-fluid-phase environment, could affect some of these sensitivities. A detailed comparison of the spatial distribution of Ti and Fe to observed SN remnants and special features of their morphology, such as the presence of ``Ni-bullets", will be the subject of future work.

\acknowledgments
This research was supported by the European Union’s Framework Programme for Research and Innovation Horizon Europe under Marie Sklodowska-Curie grant agreement No.~101065891, and by the German Research Foundation (DFG) through the Collaborative Research Centre ``Neutrinos and Dark Matter in Astro- and Particle Physics (NDM),'' Grant No.\ SFB-1258-283604770, and under Germany's Excellence Strategy through the Cluster of Excellence ORIGINS EXC-2094-390783311. The authors are grateful to the Gauss Centre for Supercomputing e.V.\ (GCS; www.gauss-centre.eu) and to the Leibniz Supercomputing Centre (LRZ; www.lrz.de) for computing time on the supercomputers SuperMUC and SuperMUC-NG at LRZ under GAUSS Call~17 and Call~20 project ID pr53yi, and to the LRZ under LRZ project ID pn69ho. They also thank the Max Planck Computing and Data Facility (MPCDF) for computing resources on the HPC systems Cobra, Draco, and Raven.

 \newpage
\bibliography{bib,ccsn,math}

\end{document}